\documentclass[%
 aip,
 amsmath,amssymb,
 reprint,%
]{revtex4-1}
\usepackage[utf8]{inputenc}
\usepackage{graphicx}
\begin{document}

\title{Hydrodynamic description of Weyl fermions in condensed state of matter}

\author{Mariya Iv. Trukhanova}
\email{trukhanova@physics.msu.ru}
 \affiliation{M.V.Lomonosov Moscow State University, Faculty of Physics, Leninskie Gory, 119991 Moscow, Russia}
\affiliation{Theoretical Physics Laboratory, Nuclear Safety Institute,
Russian Academy of Sciences, B.Tulskaya 52, 115191 Moscow, Russia}
\author{Pavel Andreev}
\affiliation{M.V.Lomonosov Moscow State University, Faculty of Physics, Leninskie Gory, 119991 Moscow, Russia}

\begin{abstract}Due to the many unique transport properties of Weyl semimetals, they are promising materials for modern electronics. We investigate the electrons in the strong coupling approximation near Weyl points based on their representation as massless Weyl fermions. We have constructed a new fluid model based on the many-particle quantum hydrodynamics method to describe the behavior of electrons gas with different chirality near Weyl points in the low-energy limit in the external electromagnetic fields, based on the many-particle Weyl equation and many-particle wave function. The derived system of equations forms a closed apparatus for describing the dynamics of the electron current, spin density and spin current density.  Based on the proposed model, we considered small perturbations in the Weyl fermion system in an external uniform magnetic field and predicted the new type of eigenwaves in the  systems of the electrons near the Weyl points.\end{abstract}

\maketitle
\tableofcontents

\section{Introduction}
\label{section1}
Weyl fermions are stable and chiral massless particles, both a particle and an antiparticle \cite{1}. Weyl fermions can be realized as emergent low-energy quasiparticle excitations in the condensed state system or Weyl semimetals \cite{2}, \cite{3}.  Semimetals, are the crystals, the conductivity of which is determined by the topological properties of the wave function in the bulk of the sample of semimetal \cite{4}, \cite{5}. Recently, these materials were discovered experimentally \cite{51}.  In the strong-coupling approximation, strongly interacting electrons can jump from one atom to another, and can be imaged
by the weakly interacting electron quasiparticles \cite{52}. In semimetals, this corresponds
to the formation of the spatial electronic band structure, where the Weyl-like points would appear in the
band structure if two non-degenerate bands cross
each other at a single node. Near this
point the quasiparticles are massless and can be described
by the Hamiltonian for the Weyl fermions \cite{6}. In the Weyl semimetal, it is precisely three-dimensional chiral charge carriers with a linear spectrum near the Weyl node, which are more stable than two-dimensional ones in graphene. Therefore, in the Weyl semimetal, the anomalous Hall effect, the presence of negative magnetoresistance, and a chiral magnetic effect are possible \cite{7}, \cite{8}.  

The study of Weyl fermions is an important and topical problem in the physics of the condensed state of matter. In addition to bulk states, charge particles
in Weyl semimetals have the gapless surface states \cite{9}. The existence of the surface
states such as Fermi arcs leads to the strong
nonlocality of transport in thin-layer
samples and  to the anomalous quantum
oscillations \cite{10}.    The Weyl semimetals exhibit the important properties in an external magnetic  and electric fields. In Ref. \cite{11} the Landau levels and quantum oscillations of the density of states in the Weyl semimetal in crossed magnetic and electric fields had been investigated and  an expression for the energy
spectrum of such system had been derived.

The transport processes which are specific to an ideal gas
of relativistic Weyl fermions had been studied in Ref. \cite{12}. The semiclassical kinetic theory of Weyl particles in the presence of external electromagnetic fields in an uniformly rotating coordinate frame, by keeping the usually ignored centrifugal force terms, had been derived \cite{13}. The phase-space dynamics of the Weyl and Dirac
particles is directly linked.  A hydrodynamic form of the Weyl equation for the neutrino wave function had been
derived in Ref. \cite{14}. A hydrodynamic field theory of the three-dimensional fractional quantum Hall effect, which was
recently proposed to exist in magnetic Weyl semimetals, when the Weyl nodes are gapped by strong repulsive
interactions, had been developed in Ref. \cite{141}. 

Weyl fermions have been actively studied in the context of superfluid $^3$He. The Weyl points can exist in the fermionic spectrum of
superfluid $^3$He-A and in the core of the quantized vortices in $^3$He-B \cite{15}. The effective or synthetic "electric" and
"magnetic" fields can be exist and act on the Weyl quasiparticles in the  topological semiconductors, because the position of the Weyl points in momentum space depends on the coordinates of space. The orbital-motive force in superfluid $^3$He-A originates from chiral Weyl fermions had been derived in Ref. \cite{16}. Tetrad formalism for the Weyl fermions and the chiral/axial anomaly in superfluid 3He-A had been discussed in Ref. \cite{17}.

The unusual properties of the low-energy quasiparticle excitations can be described by the methods ranging from the quantum field theory to the chiral kinetic and hydrodynamics theory. A steady-state nonlocal response in Weyl semimetals in the hydrodynamic regime had been studied by using the consistent hydrodynamic formalism \cite{18}. In the Drude model, which describe the electrons transport, electron scattering on impurities and phonons dominates over the electron-electron scattering. But, if electron-electron scattering prevails over electron-impurity and electron-phonon scattering, charged particles form a fluid, the properties of which can be described in terms of hydrodynamic formalism.  Also, in the hydrodynamic regime, the momentum and energy-current
relaxations are independent processes. An additional condition for the usage of chiral hydrodynamics is the smallness of the mean free path in comparison with the wavelength $k<<1/L_{mfp}$. The charge current is determined not only by the local electromagnetic field and temperature gradient, but also by the hydrodynamic flow velocity, which is described by the momentum balance equation. A theory of thermoelectric transport in weakly disordered Weyl semimetals, where
the electron-electron scattering time is faster than the electron-impurity scattering time, had been derived based on the hydrodynamical description of the relativistic fluids at each Weyl node \cite{19}. The chiral kinetic theory \cite{20}, \cite{21} and chiral hydrodynamics were constructed within the theoretical description \cite{22}, \cite{23}.  

The dynamics of a system of electrons in the strong coupling approximation can be modeled by a viscous fluid flow, and the hydrodynamic approach can be applied \cite{24}. The experimental signatures of 
hydrodynamic electron flow in the Weyl semimetal tungsten diphosphide (WP2) had been observed resently \cite{24}, using thermal and magneto-electric transport experiments. The transition 
from a conventional metallic state at higher temperatures to a hydrodynamic electron fluid 
below 20 K had been found. Thus, an important task is to describe the collective effects in Weyl semimetals based on the theory of hydrodynamics. Attempts have been made to use the model of classical hydrodynamics to describe the behavior of electrons in the strong coupling approximation \cite{25}, \cite{26}. 

On the other hand, as wave
processes, processes of information transfer, diffusion and
other transport processes occur in the three-dimensional
physical space, a need arises to turn to a mathematical
method of physically observable values which are determined
in a 3D physical space. A quantum mechanics description for the systems of $N$
interacting particles is based upon the many-particle
Schr\"{o}dinger equation (MPSE) that specifies a wave function
in a 3N-dimensional configuration space.  To do so we should derive
equations those determine dynamics of functions of
three variables, starting from MPSE. This problem has
been solved with the creation of a method of many-particle
quantum hydrodynamics (MPQHD). Therefore, the reason
for development such method is analogous with the motivation
of the density functional theory. 

In our research
we propose a further development of the MPQHD method for the hydrodynamic description of Weyl fermions in the condensed state of matter. We assume, that the electron-electron interactions dominate over the electron-impurity and electron-phonon interactions. We demonstrate the original derivation of the system of equations of hydrodynamic type. The developed model can be used in the future to study the transport of electrons in various types of Weyl semimetals.

\section{Quantum hydrodynamics formulation of Weyl equation}
The quasiparticle states in Weyl semimetals
are described by a Weyl equation. The many-particle Weyl equation can be introduce for the massless spin-$1/2$ particles in the form
\begin{equation} \label{Weyl} i\hslash\frac{\partial \psi}{ \partial t}=\sum_{j=1}^N\biggl(\lambda_j\upsilon_{f "\alpha"}(\hat{\sigma}^{\alpha}_j\hat{D}^{\alpha}_j)+\hat{\sigma}^0q_j\varphi_j\biggr)\psi,
      \end{equation}
      where $q_j$ is a charge of $j$-th particle, $\lambda_j=\pm 1$ is the chirality of $j$-th particle,
      $\upsilon_{f "\alpha"}$ is a Fermi velocity, which is the same for all electrons of the system (near the Weyl points).
      The Fermi velocity is not a vector, but a set of three parameters (two of them coincide each other).
      They are the inplane Fermi velocity $\upsilon_{f \perp}=\upsilon_{f "x"}=\upsilon_{f "y"}$,
      and the Fermi velocity $\upsilon_{f \parallel}=\upsilon_{f "z"}$ parallel to the anisotropy axis.
      Hence, the first term on the right-hand side of equation (\ref{Weyl}) can be presented in more detailed form:
      $\lambda_j\upsilon_{f "\alpha"}(\hat{\sigma}^{\alpha}_j\hat{D}^{\alpha}_j)$
      $=\lambda_j[\upsilon_{f \perp}(\hat{\sigma}^{x}_j\hat{D}^{x}_j)+\hat{\sigma}^{y}_j\hat{D}^{y}_j)
      +\upsilon_{f \parallel}\hat{\sigma}^{z}_j\hat{D}^{z}_j)]$.
      Hence, the value of parameter $\upsilon_{f "\alpha"}$ is bound to the projection of the momentum.
      It is bound to other functions below appearing via the momentum contribution.
      The difference in the Fermi velocity for different directions of space reflects the presence of anisotropy near the Weyl points. The operator   of covariant derivative  is $\hat{D}^{\alpha}_j=\hat{p}^{\alpha}_j-\frac{q_j}{c}A^{\alpha}_j$, where $\hat{p}^{\alpha}_j$ is the momentum operator near the Weyl node, $\varphi_j$ and $A^{\alpha}_j$ are the scalar and vector potentials respectively,  $\psi(R,t)=\psi(\mathbf{r}_1, \mathbf{r}_2,...,\mathbf{r}_N,t)$ is the many-particle wave function of the particles with right and left helicity  in $3N$ configuration space, and $a,b,d$ is the spin index $(=-1/2,1/2)$. The wave function has the complex form of matrix.  The Pauli matrices $\sigma^{\alpha}_j$ in the Weyl  equation  have the well known form
      $$ \qquad (\sigma^x_j)_{ab}=\begin{pmatrix}0 & 1 \\ 1 & 0 \end{pmatrix}, \qquad (\sigma^y_j)_{ab}=\begin{pmatrix}0 & -i \\ i & 0 \end{pmatrix}, $$
      \begin{equation} \qquad\qquad (\sigma^z_j)_{ab}=\begin{pmatrix}1 & 0 \\ 0 & -1 \end{pmatrix}, \qquad (\sigma^0_j)_{ab}=\begin{pmatrix}1 & 0 \\ 0 & 1 \end{pmatrix}.\end{equation}
The main idea of the many-particle quantum hydrodynamics method
\cite{Maksimov QHM 99}, \cite{Andreev PTEP 19}, \cite{Koide PRC 13}, \cite{Andreev PRE 15 SEAW}, \cite{Andreev PRB 11}, \cite{Andreev Chaos 21},
which is used in this paper, is to average the operators of observable physical quantities over quantum states.  Below, averaging of the
quantity $\chi(R,t)$ over the quantum states of particles is given by an expression of the form \begin{equation}\int dR \sum_{j=1}^N\delta(\mathbf{r}-\mathbf{r}_j)\chi(R,t)=\biggl\langle\chi(R,t)\biggr\rangle.\end{equation}
The state of the spinning massless particles is characterized by the total density in the neighborhood of $\mathbf{r}$ in a physical space. The total concentration of the particles $n(\mathbf{r},t)$ is thus determined as the quantum average of the concentration operator $\hat{n}=\sum_j\delta(\mathbf{r}-\mathbf{r}_j)$ in the coordinate representation
\begin{equation} \label{n1}
  n(\mathbf{r},t)=\biggl\langle\psi^{\dag}(R,t)\psi(R,t)\biggr\rangle=n_{+}+n_{-},
\end{equation}
where $^{\dag}$ is the Hermitian conjugation.
 Lets introduce the additional scalar physical quantity $\nu(\mathbf{r},t)$ depending on the chirality of the particles and representing the difference in the concentrations of massless particles with different chiralities
 \begin{equation} \label{n2}
  \nu(\mathbf{r},t)=\biggl\langle\lambda_j\psi^{\dag}(R,t)\psi(R,t)\biggr\rangle=n_{+}-n_{-}.
\end{equation}
 Differentiating the expressions for the total concentration (\ref{n1}) with respect to time and using the Weyl equation (\ref{Weyl}), the continuity equation for particle density $n$ can be derived in the form
\begin{equation}\label{continuity1}
  \partial_t n(\mathbf{r},t)+\upsilon_{f "\alpha"}\partial_{\alpha}j^{\alpha}(\mathbf{r},t)=0,
\end{equation}
where the analog of mass current in the continuity equation in the fluid hydrodynamics is derived in the form of the vector physical quantity, which represents the difference between spin densities of particles with different chiralities  \begin{equation}\label{current}   j^{\alpha}(\mathbf{r},t)=\biggl\langle\lambda_j\psi^{\dag}(R,t)\sigma^{\alpha}_j\psi(R,t)\biggr\rangle=s^{\alpha}_{+}-s^{\alpha}_{-}.  \end{equation}    This result follows from the fact that the velocity of massless particles is connect with spin, which is codirectional or opposite to the direction of motion of the particle.
Symbol $\upsilon_{f "\alpha"}$ in the continuity equation (\ref{continuity1}) is not a vector.
Subindex $"\alpha"$ refers to the one of two values of this parameter,
which are bound to the value of the vector index $\alpha$ existing in the same term.

Continuity equation for the quantity $\nu(\mathbf{r},t)$ can be derived in a similar way in the form
 \begin{equation}\label{continuity2}
  \partial_t \nu(\mathbf{r},t)+\upsilon_{f "\alpha"}\partial_{\alpha}s^{\alpha}(\mathbf{r},t)=0,
\end{equation}
where  the divergence of the spin density vector appears during the derivation of the equation (\ref{continuity2}) in the microscopic form
\begin{equation}
\label{spin}   s^{\alpha}(\mathbf{r},t)=\biggl\langle\psi^{\dag}(R,t)\sigma^{\alpha}_j\psi(R,t)\biggr\rangle=s^{\alpha}_{+}+s^{\alpha}_{-}.  \end{equation} The important properties of the Weyl fermions is the existence of chiral anomaly, when for the fermions with different chiralities the currents 
of the left-handed and right-handed Weyl fermions cannot be preserved separately. But within the framework of the method of quantum hydrodynamics, the equation of dynamics for the $\nu(\mathbf{r},t)$ has the form of a conservation law.

It is of interest to obtain a closed apparatus of quantum hydrodynamics, consisting of a system of equations for the basic physical quantities. For this purpose we need to follow the dynamics of vectors $s^{\alpha}(\mathbf{r},t)$ and $j^{\alpha}(\mathbf{r},t)$ in time. By repeating the above procedure for spin density vector $s^{\alpha}$ and vector $j^{\alpha}$, the balance equations for these quantities can be derived in the way \begin{equation}\label{spin equation}  \partial_t s^{\alpha}(\mathbf{r},t)+\upsilon_{f "\alpha"}\partial_{\alpha}\nu(\mathbf{r},t)=\frac{2}{\hslash}
\upsilon_{f "\beta"}\varepsilon^{\alpha\beta\gamma}\tau^{\gamma\beta}(\mathbf{r},t)   \end{equation}          and \begin{equation}\label{current equation}  \partial_t j^{\alpha}(\mathbf{r},t)+\upsilon_{f "\alpha"}\partial_{\alpha}n(\mathbf{r},t)=\frac{2}{\hslash}
\upsilon_{f "\beta"}\varepsilon^{\alpha\beta\gamma}\Lambda^{\gamma\beta}(\mathbf{r},t).   \end{equation}
As we can see the balance equations (\ref{spin equation}) and (\ref{current equation}) do not include the effect of the fields of external forces. But tensors of the second rank $\tau^{\gamma\beta}$, $\Lambda^{\gamma\beta}$ appear at the right-hand side of the equations (\ref{spin equation}) and (\ref{current equation}), the physical meaning of which requires clarification.  The terms at the right hand sides of the equations  characterize the new force fields act on the spin densities of the particles with difference chiralities. In the process of derivation of the system of equations (\ref{spin equation}) and (\ref{current equation}) tensors $\tau^{\gamma\beta}$ and $\Lambda^{\gamma\beta}$ can be obtained in the microscopic form
     $$ \tau^{\alpha\beta}(\mathbf{r},t)
       =\biggl\langle\frac{\lambda_j}{2}\biggl(\psi^{\dag}(R,t)\sigma^{\alpha}_jD^{\beta}_j\psi(R,t)$$ \begin{equation}\label{tau}\qquad\qquad\qquad\qquad+(D^{\beta}_j\psi)^{\dag}(R,t)\sigma^{\alpha}_j\psi(R,t)\biggr)\biggr\rangle\end{equation}
and $$\Lambda^{\alpha\beta}(\mathbf{r},t)
       =\biggl\langle\frac{1}{2}\biggl(\psi^{\dag}(R,t)\sigma^{\alpha}_jD^{\beta}_j\psi(R,t)$$\begin{equation}\label{lyamda}
       \qquad\qquad\qquad\qquad+(D^{\beta}_j\psi)^{\dag}(R,t)\sigma^{\alpha}_j\psi(R,t)\biggr)\biggr\rangle.\end{equation}
 As we can see, the quantity $\Lambda^{\alpha\beta}$ represents the total spin current density of the massless fluid of particles.  Now it is important to derive the evolution equations for the spin current density tensor $\Lambda^{\alpha\beta}$ and the tensor $\tau^{\alpha\beta}$, which characterizes the difference of spin currents of the particles with different chiralities. Differentiating tensor quantities (\ref{tau}) and (\ref{lyamda}) with respect to time and using the Weyl equation (\ref{Weyl}), we arrive at the system of equations for the evolution of the spin current density
   $$\partial_t\tau^{\alpha\beta}+\upsilon_{f " \alpha"}\partial^{\alpha}J^{\beta}+\frac{\hslash}{2}\upsilon_{f " \delta"}\varepsilon^{\alpha\gamma\delta}\partial_{\beta}\partial_{\delta}s^{\gamma}= qj^{\alpha}E^{\beta}$$
   \begin{equation}  \label{j1}
   -\frac{q}{c}\upsilon_{f " \alpha"}\varepsilon^{\alpha\beta\gamma}nB^{\gamma}+\frac{2}{\hslash}\varepsilon^{\alpha\gamma\delta}\upsilon_{f "\gamma"}p^{\delta\beta\gamma},  \end{equation}
   $$\partial_t\Lambda^{\alpha\beta}+\upsilon_{f " \alpha"}\partial^{\alpha}\tau^{\beta}+\frac{\hslash}{2}\upsilon_{f " \delta"}\varepsilon^{\alpha\gamma\delta}\partial_{\beta}\partial_{\delta}j^{\gamma}= qs^{\alpha}E^{\beta}$$ \begin{equation}  \label{j2}
   -\frac{q}{c}\upsilon_{f " \alpha"}\varepsilon^{\alpha\beta\gamma}\nu B^{\gamma}+\frac{2}{\hslash}\varepsilon^{\alpha\gamma\delta}\upsilon_{f "\gamma"}\Upsilon^{\delta\beta\gamma}.  \end{equation}
The second terms at the left hand side of equations (\ref{j1}) and (\ref{j2}) represent the influence of the nonuniform total current $J^{\beta}$ and axial current $\tau^{\beta}$ on the dynamics of spin current densities of the particles. In the similar way the third terms at the left-hand side of equations represent the influence of nonuniform spin density fields. The first and second terms at the right-hand sides represent the influence of external electric and magnetic fields. As we can see, the first term at the right-hand sides are the influence of external electric field on the spin densities of the particles with different chiralities. The second terms  at the right-hand sides represent the effects of external magnetic field on the particles densities. External electric and magnetic fields can be uniform. And the last terms at the right hand sides of equations (\ref{j1}) and (\ref{j2})  are the tensors of the third rank. The second terms at the left hand side of equations associated with the influence of an inhomogeneous total current density $J^{\beta}$
 \begin{equation}J^{\alpha}(\mathbf{r},t)
       =\biggl\langle\frac{1}{2}\biggl(\psi^{\dag}(R,t)D^{\alpha}_j\psi(R,t)+h.c.\biggr)\biggr\rangle=J^{\alpha}_{+}+J^{\alpha}_{-}\end{equation}
   and the axial current density
\begin{equation}\tau^{\alpha}(\mathbf{r},t)
       =\biggl\langle\frac{\lambda_j}{2}\biggl(\psi^{\dag}(R,t)D^{\alpha}_j\psi(R,t)+h.c.\biggr\rangle=J^{\alpha}_{+}-J^{\alpha}_{-}\end{equation}
At the last stage, an important task is to derive the equations for the dynamics of the total current density $J^{\alpha}$ representing the sum of partial currents and axial current density $\tau^{\alpha}$ representing the difference in partial currents with different chiralities. Repeating the above-described differentiation procedure, we arrive at equations of the form
 \begin{equation}\label{j3}\partial_tJ^{\alpha}+\upsilon_{f " \beta"}\partial_{\beta}\tau^{\beta\alpha}=qnE^{\alpha}+\frac{q}{c}\upsilon_{f " \beta"}\varepsilon^{\alpha\beta\gamma}j^{\beta}B^{\gamma}\end{equation}
   and    \begin{equation}\label{j4}\partial_t\tau^{\alpha}+\upsilon_{f " \beta"}\partial_{\beta}\Lambda^{\beta\alpha}=q\nu E^{\alpha}+\frac{q}{c}\upsilon_{f " \beta"}\varepsilon^{\alpha\beta\gamma}s^{\beta}B^{\gamma}.\end{equation}
 The first equation (\ref{j3}) is the momentum balance equation for the total current density. The force field at the right hand side of equation  represents the density of the  force acting on a charge density and spin density in an external electromagnetic field. The second equation (\ref{j4}) is the equation for the evolution of the axial current density $\tau^{\alpha}$.   The second term at the right hand side of equation for the axial current density (\ref{j4}) represents the action of torque in an external magnetic field.

In external uniform magnetic and electric fields, the system of equations (\ref{continuity1}), (\ref{continuity2}), (\ref{spin equation}), (\ref{current equation}), (\ref{j1}), (\ref{j2}), (\ref{j3}) and (\ref{j4}) should lead to known effects, such as a linear spectrum in a magnetic field. In addition, the developed model should predict the appearance of new effects that are absent in the one-particle description.
 \section{Hydrodynamic of Weyl fermions with single chirality}
 Let us simplify the problem and consider the one-helical state.
Dual structure of the hydrodynamic equations obtained above related to the interplay of two values of chirality.
If we consider the evolution of the system of particle with positive chirality
we find simplified set of four hydrodynamic equations.
The continuity equation has no changes in compare with its original form (\ref{continuity1})
\begin{equation}\label{continuity 1 sp}
\partial_t n+\upsilon_{f "\alpha"}\partial_{\alpha}j^{\alpha}=0.
\end{equation}
The current of particles $j^{\alpha}$ leads to the equation of kinematic nature
\begin{equation}\label{current equation 1 sp}  \partial_t j^{\alpha}+\upsilon_{f "\alpha"}\partial_{\alpha}n=\frac{2}{\hslash}
\upsilon_{f "\beta"}\varepsilon^{\alpha\beta\gamma}\Lambda^{\gamma\beta}.\end{equation}
No interaction contributes in equation (\ref{current equation 1 sp}),
but additional function $\Lambda^{\gamma\beta}$ enters it.
It shows the way to extend the set of hydrodynamic equations to include the action of the electromagnetic field on the collection of the Weyl fermions.
The evolution of tensor $\Lambda^{\alpha\beta}$ connected to the dynamic of the momentum density $J^{\alpha}$.
These equations have the following form
which is simplified in compare with the similar equations obtained above in the general regime
\begin{equation}\label{equation for J alpfa 1 sp}
\partial_t J^{\alpha}+\upsilon_{f "\beta"}\partial_{\beta}\Lambda^{\beta\alpha}=qnE^{\alpha}
+\frac{q}{c}\upsilon_{f "\beta"}\varepsilon^{\alpha\beta\gamma}j^{\beta}B^{\gamma},\end{equation}

$$\partial_t\Lambda^{\alpha\beta}+\upsilon_{f "\alpha"}\partial^{\alpha}J^{\beta}
+\frac{\hslash}{2}\upsilon_{f "\delta"}\varepsilon^{\alpha\gamma\delta}\partial_{\beta}\partial_{\delta}j^{\gamma}$$
\begin{equation}  \label{equation for Lambda 1 sp}\qquad
=qs^{\alpha}E^{\beta}-\frac{q}{c}\upsilon_{f "\alpha"}\varepsilon^{\alpha\beta\gamma}n B^{\gamma}
+\frac{2}{\hslash}\varepsilon^{\alpha\gamma\delta}\upsilon_{f "\gamma"}\Upsilon^{\delta\beta\gamma}.  \end{equation}
We have a closed system of equations that determine the dynamics of massless charged Weyl particles against the background of external electromagnetic fields. The equation (\ref{equation for J alpfa 1 sp}) is the momentum balance equation, according to which the dynamics of the flow velocity is affected by the electric field acting on the particle number density and the magnetic field acting on the spin density. The equation (\ref{equation for Lambda 1 sp}) represents the spin current density evolution equation. As we can see from the right hand side of equation, the evolution of the spin current is associated with the action of an external electric field on the spin density and with the action of an external magnetic field on the concentration of the number of particles. The effect may have both homogeneous and not-uniform electromagnetic fields.

\section{Wave propagation in system of Weyl fermions with positive chirality}

One of the primary goals of this article is to derive dispersion
characteristics of eigenwaves in the  systems of the electrons near the Weyl points
in the external uniform magnetic field. In this section we consider the systems
of charged Weyl particles with spin aligned to the velocity $\lambda=+1$, $q=-e$.  The external magnetic field aligned with $z$ direction $\mathbf{B}=\{0,0,B_0\}$. The wave propagates in the parallel direction $\mathbf{k}=\{0,0,k_{\parallel}\}$. For the Fermi velocity  To do that let’s analyze small perturbations
of physical variables from the stationary state
$$n=n_0+\delta n, \qquad j^{\alpha}=0+\delta j^{\alpha},$$
\begin{equation} J^{\alpha}=0+\delta J^{\alpha},\qquad \Lambda^{\alpha\beta}=0+\delta \Lambda^{\alpha\beta}.\end{equation}
Considering small deviations from the equilibrium position in the equations for the particle number density, spin density, and spin current density, we arrive at the two dispersion equations for the longitudinal wave
  \begin{equation}\label{w}-\omega^2+k_{\parallel}^2\upsilon^2_{f \parallel}=-\frac{4}{\hslash}\upsilon^2_{f\perp}\frac{eB_0}{c}\frac{k_{\parallel}\upsilon_{f \parallel}}{\omega}\end{equation} and transversal wave
  \begin{equation}\label{w1} \omega^4-k^4_{\parallel}\upsilon^4_{f \parallel}=\frac{2}{\hslash}\upsilon^2_{f\perp}\frac{eB_0}{c}k_{\parallel}\upsilon_{f \parallel}\omega. \end{equation}
      In the limit of low magnetic fields, the frequency spectrum (\ref{w}) is characterized by a dispersion relation for a stable wave
      \begin{equation}\label{w3}\omega=\sqrt{k_{\parallel}^2\upsilon^2_{f \parallel}+\frac{4}{\hslash}\upsilon^2_{f\perp}\frac{eB_0}{c}},\end{equation}
  as can be seen from the dispersion law, the frequency for the Weyl fermion system in an external magnetic field $B_0$ is shifted relative to the frequency of eigenwaves \begin{equation}\label{w4}\omega=\upsilon_{f \parallel}k_{\parallel}\end{equation} in the absence of external field, when the wave propagates at the Fermi velocity $\upsilon_{f \parallel}$. The linear spectrum is derived for the positive chirality.
  
  In strong magnetic fields, an unstable wave solution arises, which is very predictable, since the magnetic field, directed perpendicular to the direction of wave propagation, tends to rotate the Weyl fermions in the plane $yz$.
  In the regime of long wave length the dispersion relation (\ref{w}) takes the form
  \begin{equation}\label{w2}\omega=\sqrt[3]{\frac{4}{\hslash}\upsilon^2_{f\perp}\frac{eB_0}{c}k_{\parallel}\upsilon_{f "\parallel"}}. \end{equation}
Equation (\ref{w2}) predicts an unexpected result for Weyl fermions in strong magnetic fields. 

The law of dispersion for the low magnetic fields (\ref{w3}) is nonlinear, which is very specific. The spectrum of one-electron states in a magnetic field represents the Landau levels. The zero Landau level has linear dispersion, the sign of which depends on the state of chirality. In the absence of an external magnetic field, the dispersion law (\ref{w4}) coincides with the linear spectrum of the zero Landau level, as expected. But, magnetic field strongly changes the spectrum making it nonlinear. The nonlinear behavior of the dispersion laws may be related to the fact that we considered the collective dynamics of a system of a large number of Weyl fermions, while the Hamiltonian of the one-particle model gives Landau zones.

\section{Conclusion}
Weyl fermions have long been a hypothetical particle, but with the discovery of Weyl semimetals, Weyl fermions were found in the form of quasiparticles.
Weyl fermions are highly mobile and can overcome obstacles that slow down ordinary electrons. They are not scattered by crystal defects, their movement is stable and not supported by the influence of noise. Due to their unique properties, fermions can be used to create fast electronic devices.

Therefore, an important task is the development of theoretical models for describing the behavior of electrons in the state of Weyl fermions. In this article, based on the method of many-particle quantum hydrodynamics, we consider the hydrodynamical model of a fluid of massless particles with charge in the external electromagnetic fields. It is assumed that the electron-electron scattering prevails over the electron-impurity and electron-phonon scattering, and charged particles form a fluid, the properties of which can be described in terms of hydrodynamic formalism.   Based on the many-particle Weyl equation (\ref{Weyl}) and many-particle wave function $\psi(R,t)$, a closed system of equations of hydrodynamic type was derived, which consists of eight equations: the equations of continuity (\ref{continuity1}) and (\ref{continuity2}),  spin density evolution equations (\ref{spin equation}), (\ref{current equation}), spin currents density evolution equations (\ref{j1}) and (\ref{j2}), momentum balance equations for the total current and axial current densities (\ref{j3}), (\ref{j4}). The equations for the balance of the spin current and the balance of the momentum contain force fields representing the action of the external electromagnetic fields. The constructed hydrodynamic model for describing chiral charge carriers in external fields differs from the previously developed models of classical hydrodynamics for describing Weyl fermions \cite{19}, \cite{22}, \cite{23}.  

We used a new mathematical model to study the wave dispersion in a  system of the Weyl particles and predicted the new type of eigenwaves in the  systems of the electrons near the Weyl nodes
in the external uniform magnetic field, which is directed along the $"z"$ axis. For the case of the longitudinal and transversal waves (\ref{w}) and (\ref{w1}) in the absence of an external magnetic field, the dispersion law characterizes the linear dependence of the wave frequency on the wave number when the wave propagates at the Fermi velocity. The magnetic field strongly changes the dispersion laws.

A new theoretical model for the description of the Weyl fermion gas based on the fluid model of many-particle quantum hydrodynamics is obtained in the article. A further task for the development of this method is to study the effect of an electric field on the Landau zones, on the chiral properties of Weyl semimetals, and also on surface states in the form of Fermi arcs. Moreover, further development of the model requires taking into account the interactions between Weyl fermions.

\begin{acknowledgements}
The work of Trukhanova Mariya Iv. is supported by the Russian Science Foundation  under grant $No.$ $19-72-00017$.
The contribution of Pavel Andreev in this paper has been supported by the RUDN University Strategic Academic Leadership Program.
\end{acknowledgements}



\begin{thebibliography}{9}

\bibitem{1} H. Weyl, Z. Phys., {\bf 56}, 330(1929).
\bibitem{2} H. Weng, et al., Phys. Rev. X, {\bf 5}, 011029 (2015).
\bibitem{3} S. M. Huang, et al., Nat. Commun., {\bf 6}, 7373 (2015).
\bibitem{4} Sh. Jia, S. Y. Xu, and M. Z. Hasan, Nature Mater., {\bf 15}, 1140 (2016).
\bibitem{5} S. Y. Xu et al., Science, {\bf 349}, 613 (2015).
\bibitem{51} I. Belopolski, D. S. Sanchez et al., Nature Commun., {\bf 7}, 13643 (2016).
\bibitem{52} A. R. Battye and A. Moss, Phys. Rev. Lett., {\bf 112}, 051303 (2014).
\bibitem{6} Shengyuan A. Yang, SPIN, {\bf 6}, No 1640003, 1 (2016).
\bibitem{7} S. A. Parameswaran, et al., Phys. Rev. X, {\bf 4}, 031035 (2014).
\bibitem{8} D. Bulmash and X.-L. Qi, Phys. Rev. B, {\bf 93}, 081103(R) (2016).
\bibitem{9} Y. Alavirad and J. D. Sau, Phys. Rev. B, {\bf 94}, 115160 (2016).
\bibitem{10} Y. Baum, E. Berg, S. A. Parameswaran, and A. Stern, Phys. Rev. X, {\bf 5}, 041046 (2015).
\bibitem{11} Z. Z. Alisultanov, JETP, {\bf 152}, 986 (2017).
\bibitem{12} R. Loganayagam, P. Surowka,  J. High Energ. Phys., {\bf 2012}, 97 (2012).
\bibitem{13} F. Dayi, E. Kilinzarslan and E. Yunt, Physical Review D, {\bf 95}(8), 085005 (2017).
\bibitem{14} Iwo Bialynicki-Birula, Acta Physica Polonica Series B, {\bf 26}(7) (1995).
\bibitem{141} Manisha Thakurathi and A. A. Burkov, Phys. Rev. B, {\bf 101}, 235168 (2020).
\bibitem{15} G. E. Volovik,  JETP Letters, {\bf 103:2}, 140 (2016).
\bibitem{16} G. E. Volovik, JETP Letters, {\bf 98}(8), 480 (2013).
\bibitem{17} J. Nissinen, G.E. Volovik, JETP, {\bf 127}(5), 948 (2018).
\bibitem{18} E. V. Gorbar, V.A. Miransky, I.A. Shovkovy, P.O. Sukhachov, Phys. Rev., B, {\bf 98}, 035121 (2018).
\bibitem{19} Andrew Lucas, Richard A. Davison, Subir Sachdev, Proceedings of the National Academy of Sciences, {\bf 113}, 9463 (2016).
\bibitem{20} D. T. Son and N. Yamamoto, Phys. Rev. Lett., {\bf 109}, 181602 (2012).
\bibitem{21} J. Y. Chen, D. T. Son, M. A. Stephanov, H. U. Yee, and Y. Yin, Phys. Rev. Lett., {\bf 113}, 182302 (2014).
\bibitem{22} Y. Hidaka, S. Pu, and D. L. Yang, Phys. Rev. D, {\bf 97}, 016004 (2018).
\bibitem{23} Y. Neiman and Y. Oz, JHEP, {\bf 1103}, 023 (2011).
\bibitem{24} Gooth, J., Menges, F., Kumar, N. et al., Nat Commun., {\bf 9}, 4093 (2018).
\bibitem{25} A. Lucas, J. Crossno, K. C. Fong, P. Kim and S. Sachdev, Phys. Rev. B - Condens. Matter Mater. Phys., {\bf 93}, (2016).
\bibitem{26} A. Principi and G. Vignale,   Phys. Rev. Lett., {\bf 115}, (2015).
\bibitem{Maksimov QHM 99} L. S. Kuz'menkov, S. G. Maksimov, Theor. Math. Phys., {\bf 118}, 227 (1999).
\bibitem{Andreev PTEP 19} P. A. Andreev, L. S. Kuz'menkov, Prog. Theor. Exp. Phys., {\bf 2019}, 053J01 (2019).
\bibitem{Andreev PRE 15 SEAW} P. A. Andreev, Phys. Rev. E, {\bf 91}, 033111 (2015).
\bibitem{Koide PRC 13}T. Koide, Phys. Rev. C, {\bf 87}, 034902 (2013).
\bibitem{Andreev PRB 11}P. A. Andreev, L. S. Kuzmenkov, M. I. Trukhanova, Phys. Rev. B, {\bf 84}, 245401 (2011).
\bibitem{Andreev Chaos 21}P. A. Andreev, Chaos, {\bf 31}, 023120 (2021).

\end{thebibliography}
\end{document}